\documentclass[journal=nalefd,manuscript=letter,layout=twocolumn,nofootinbib]{achemso}

\usepackage{achemso}
\usepackage[latin1]{inputenc}
\usepackage[english]{babel}
\usepackage{amsmath}
\usepackage{amsfonts}
\usepackage{amssymb}
\usepackage{graphicx}
\usepackage{pst-all}
\usepackage{psfrag}
\usepackage{bm}
\usepackage{epsfig}
\usepackage{float}

\let\oldmaketitle\maketitle
\let\maketitle\relax

\usepackage[sort&compress,numbers,super]{natbib}

\DeclareCaptionLabelFormat{myformat}{\bf{#1~#2}}
\title{\Large \bf Phonon-mediated mid-infrared photoresponse of graphene}

\author{M.~Badioli}
\author{A.~Woessner}
\author{K.~J.~Tielrooij}
\author{S.~Nanot}
\author{G.~Navickaite}
\affiliation{ICFO -  Institut de Ci\`encies Fot\`oniques, Mediterranean Technology Park, Av. Carl Friedrich Gauss 3, 08860 Castelldefels (Barcelona), Spain}
\author{T.Stauber}
\affiliation{Departamento de Teor\'{\i}a y Simulaci\'on de
Materiales, Instituto de Ciencia de Materiales de Madrid, CSIC,
E-28049 Madrid, Spain}
\author{F.~J.~Garc\'{\i}a~de~Abajo}
\affiliation{ICFO - Institut de Ci\`encies Fot\`oniques, Mediterranean Technology Park, Av. Carl Friedrich Gauss 3, 08860 Castelldefels (Barcelona), Spain}
\alsoaffiliation{ICREA - Instituci\'o Catalana de Recerca i Estudis Avan\c{c}ats, Passeig Llu\'{\i}s Companys, 23, 08010 Barcelona, Spain}
\author{F.~H.~L.~Koppens}
\affiliation{ICFO -  Institut de Ci\`encies Fot\`oniques, Mediterranean Technology Park, Av. Carl Friedrich Gauss 3, 08860 Castelldefels (Barcelona), Spain}
\email{frank.koppens@icfo.eu}

\newcommand{\cm}{cm$^{-1}$}
\newcommand{\um}{$\mu$m}

\begin{document}

\twocolumn[
\begin{@twocolumnfalse}
\oldmaketitle
\vspace{-1.2cm}
\begin{abstract}
The photoresponse of graphene at mid-infrared frequencies is of high technological interest and is governed by fundamentally different underlying physics than the photoresponse at visible frequencies, as the energy of the photons and substrate phonons involved have comparable energies. Here we perform a spectrally resolved study of the graphene
photoresponse for mid-infrared light by measuring spatially resolved photocurrent over a broad frequency range (1000-1600 \cm). We unveil the different mechanisms that give rise to photocurrent generation in graphene on a polar substrate. In particular, we find an enhancement of the photoresponse when the light excites bulk or surface phonons of the SiO$_2$ substrate. This work paves the way for the development of graphene-based mid-infrared thermal sensing technology.
\end{abstract}
\end{@twocolumnfalse}
]
 
\noindent
Graphene photonics\cite{Bonaccorso2010a}, optoelectronics\cite{Ferrari2014} and nanophotonics\cite{Grigorenko2012a,GarciadeAbajo2014a, Low2014c} are fast growing fields with increasing attention for the mid-infrared (MIR) spectral regime. This spectral region is interesting from both fundamental and technological points of view. For example, MIR covers the characteristic vibrational frequency range of many relevant molecules, as well as most of the thermal radiation emitted from warm objects. Therefore, the MIR range is  crucial for spectroscopy and biosensing, and for thermal imaging in applications ranging from medical diagnostics to damage-assessment, and defense. In particular, graphene is a promising material for detecting MIR light because it provides additional features with respect to current technologies, such as broadband absorption for infrared and visible light, in-situ tunable carrier density, easy integration with silicon electronics, room temperature operation, and flexibility \cite{Bonaccorso2010a, Koppens2014}.\\
In addition, graphene MIR physics is enriched by the fact that the energy scale of the photons is comparable to the Fermi energy, the energy of quasi-particle excitations such as plasmons\cite{Grigorenko2012a, GarciadeAbajo2014a, Low2014c}, and the energy of intrinsic and substrate phonons. Concerning graphene-phonon interactions, extensive studies on scattering of graphene carriers by surface phonons of polar substrates in the context of electron transport properties \cite{Chen2008, Meric2008} and relaxation dynamics of hot carriers  \cite{Hwang2013} have been performed. In a related context, electron-phonon scattering at surfaces has been widely studied to understand photoemission spectra \cite{Wang1972}. For graphene, surface phonons have been found to limit the graphene mobility \cite{Chen2008, Meric2008, Fratini2008}, and to provide additional cooling pathways of photoexcited carriers \cite{Freitag2013b, Low2012}. \\
All these phenomena are based on the interaction between electrons and thermally occupied phonon states. In contrast, the use of mid-infrared light can provide an efficient way to excite bulk or surface phonons, which can in turn act back on the graphene charge carriers. Therefore, the graphene MIR photoresponse involves a subtle interplay of light-graphene and light-substrate interactions. Photocurrent measurements can thus provide a probe for MIR graphene photophysics, as well as a starting point towards photodetection applications. In particular, light-substrate interactions can be exploited to enhance the MIR detection efficiency.\\
Pioneering works have already demonstrated bolometric\cite{Yan2012a,Freitag2013} or photoconductive\cite{Yao2014} MIR detection with graphene, and a photo-thermoelectric photoresponse of graphene p-n junctions to 10.6 \um~ light has been studied \cite{Herring2014}, showing higher responsivity for the appropriate choice of substrate. In nano-patterned graphene, the bolometric photoresponse has been observed to be enhanced by the plasmon-phonon polariton supported by the substrate \cite{Freitag2014}. Furthermore, photocurrents arising from photo-galvanic and photon drag effects under oblique incidence on large-area epitaxially grown graphene have been reported \cite{Jiang2011b, Olbrich2013}, showing a photoresponse related to the substrate reflection \cite{Olbrich2013}.\\
Here we present a detailed study on the photoresponse mechanism of planar graphene for MIR light, and the role of the substrate. We show clear evidence of two distinct mechanisms that contribute to MIR photocurrent generation, obtained from spectrally resolved photocurrent measurements. The first of these is an indirect mechanism, where substrate phonons absorb light and subsequently locally heat up graphene charge carriers. This generates a photo-thermoelectric voltage as carriers of different temperatures equilibrate, which causes charge flow. The second mechanism is based on direct light absorption in graphene, resulting in hot electrons that generate photocurrent. This latter mechanism turns out to be greatly enhanced by the substrate surface phonons, which produce a strong concentration of the near electric field, and therefore mediate enhanced light absorption at MIR frequencies. We remark that all of the photoresponse enhancement mechanisms discussed in this work operate under zero source-drain bias conditions, which are promising for low dark-current MIR detection applications.\\
The experimental results presented here are based on spatially and spectrally resolved photocurrent maps. We excite the graphene transistor samples with light from a quantum cascade laser (QCL) tuned to a specific frequency in the MIR range (1000-1600 \cm). The light is focused down to a spot whose FWHM is comparable to the wavelength. The samples consist of patterned chemical vapor deposition (CVD) grown graphene contacted by two gold electrodes (source and drain) with a separation distance $\geq$ 100 \um, on top of a substrate of 300 nm thermally grown SiO$_2$ and weakly doped silicon (more details in Methods). We record the photo-induced current that flows between source and drain, without applying a bias voltage. The doped silicon back-gate allows us to control the Fermi energy through capacitive coupling. The transmitted light is re-collimated and collected, thus enabling us to perform simultaneous acquisition of the frequency-resolved graphene photoresponse and optical transmission. We measured eight different samples that exhibited the same trends, and here we present the results of three of them (see Methods).\\
In Figures 1a,b we show the spatially resolved photocurrent $I_{\rm{PC}}$, normalized by the incoming power $P_{\rm{inc}}$, for two excitation wavelengths, $\lambda$~= 9.26 \um~ and $\lambda$~= 7.19 \um, corresponding to $\tilde{\nu}_{\rm{e}}=$1080 \cm~ and $\tilde{\nu}_{\rm{e}}=$1390 \cm,  respectively. In both cases, the normalized photocurrent  $PC_{\rm{norm}}=I_{\rm{PC}}/P_{\rm{inc}}$ peaks close to the graphene/contact edge, as previously reported for visible and near-infrared light \cite{Lee2008, Park2009a, Xia2009, Mueller2009}. We note that for both frequencies the polarization of the incident light is parallel to the contact edge. Moreover, the position of the contacts is retrieved from the transmission measured at the same time and the photocurrent dependence on power is linear (see Supplementary Information).\\

\begin{figure*}[h!!t]
\includegraphics[scale=1]{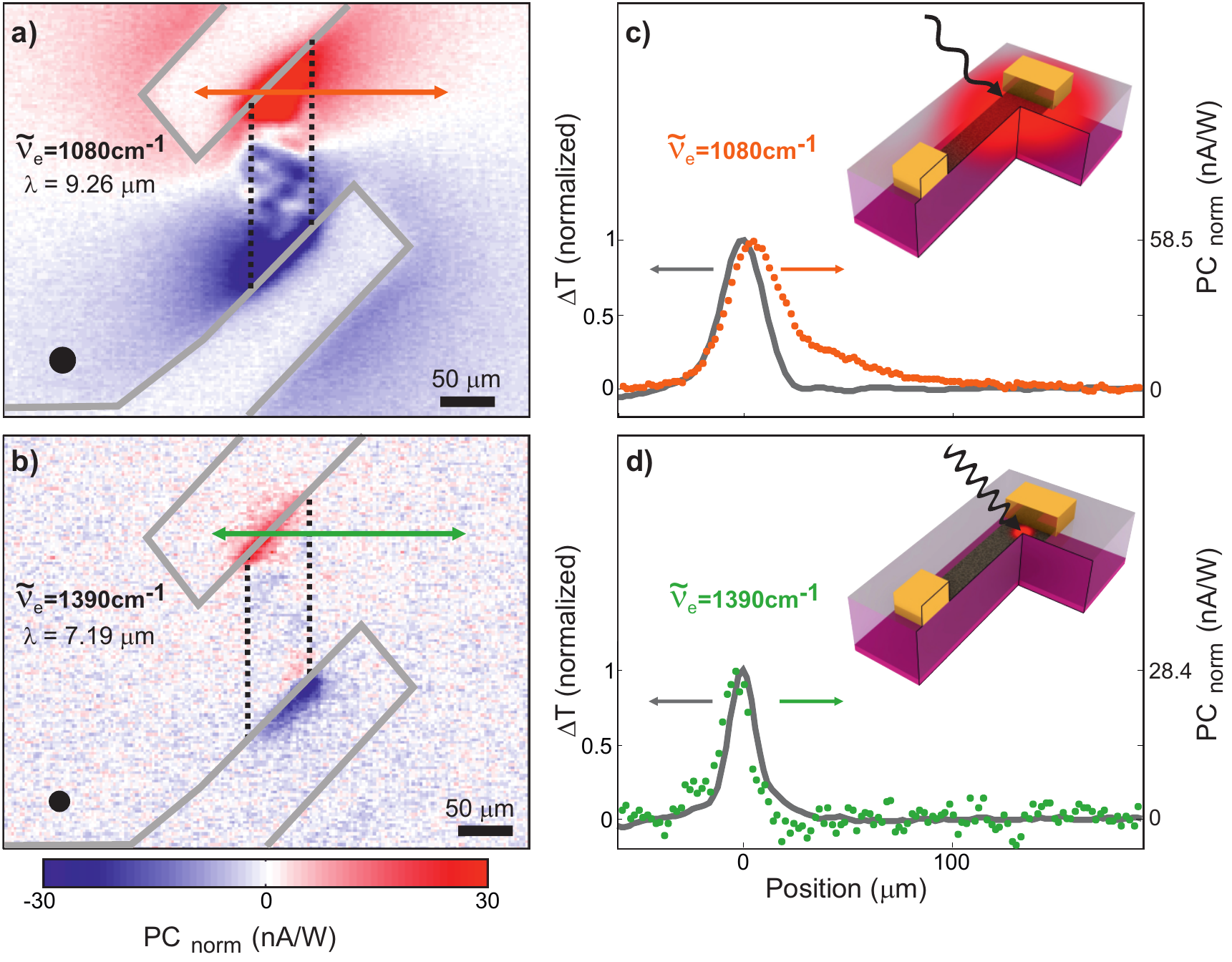}
\label{fig1}
\caption{Photocurrent maps upon excitation with wavelength $\lambda=9.26$ \um~ (1080 \cm) (a), and $\lambda=7.19$ \um~ (1390 \cm) (b). The grey lines indicate the position of the contacts retrieved from the transmission measurement; the black dotted lines indicate the graphene position. The black circle at the bottom left represents the beam spot size (FWHM 22 \um~(a) and 14 \um~(b)). The photocurrent values are normalized by the incident power. The light polarization is parallel to the contacts. (c) In orange, spatial linecut of the photocurrent along the orange arrow in (a), for 1080 \cm. (d) In green, spatial linecut of the photocurrent along the green arrow in (b), for 1390 \cm. In both cases the grey curves are the normalized spatial derivatives of the corresponding transmission measurement, which indicate the laser spot size. The sketches on the right of (c) and (d) represent artistic views of the two photocurrent generation mechanisms, due to the substrate absorption (c) and the graphene absorption (d).}
\end{figure*}
\noindent
Interestingly, there is a clear difference in the strength and spatial extent of $PC_{\rm{norm}}$ for the two frequencies. For $\tilde{\nu}_{\rm{e}}=$1080 \cm,  the normalized photocurrent is high compared to $\tilde{\nu}_{\rm{e}}=$1390 \cm. Furthermore, for $\tilde{\nu}_{\rm{e}}=1080$ \cm~ the system is photo-responsive over a much larger area that extends even outside the graphene sheet, while the response for $\tilde{\nu}_{\rm{e}}=1390$ \cm~ is confined to the interface between graphene and the electrodes. We recall that for excitation with VIS and NIR light the photoresponsive area has been shown to be between hundreds of nanometers \cite{Mueller2009} and a few microns \cite{Gabor2011}, hence much smaller than the photoresponsive area we observe here.\\
For a more quantitative comparison, we show in Figure 1c,d a cut of $PC_{\rm{norm}}$ across one of the contacts and compare it to the size of the excitation beam, which we obtain from the spatial derivative of the transmitted light. The data show that the photocurrent generated for $\tilde{\nu}_{\rm{e}}=$1390 \cm~ is spot-size limited. In contrast, for $\tilde{\nu}_{\rm{e}}=$1080 \cm~ the photoresponsive area is significantly larger than the spot-size: $\sim$20\% of the peak signal is still present 50 \um~ away from the contact edge. In addition, the overall signal at the edge is twice as large as for $\tilde{\nu}_{\rm{e}}=1390$ \cm.\\

These observations suggest the presence of different photocurrent generation mechanisms for the two wavelengths, which we aim to identify by combining spectrally resolved photoresponse and substrate transmission measurements for the $\sim$6-10 \um~ wavelength range ($\tilde{\nu}_{\rm{e}}=$ 1000-1600 \cm). The photocurrent spectrum is shown in Figure 2a. We observe a clear peak at $\tilde{\nu}_{\rm{e}}=1080$ \cm, with a shoulder extending up to $\tilde{\nu}_{\rm{e}}=1280$ \cm.  The frequency of the photocurrent peak coincides with a peak in the substrate extinction spectrum $1-T$, with $T$ being the transmission (obtained from FTIR measurements), as shown in the inset of Figure 2a. This indicates that the absorption in the substrate has a strong effect on the photocurrent.\\
The absorption in the SiO$_2$ (see Figure 2b) is dominated by two bulk optical phonon modes within our frequency window: transverse (TO) and longitudinal (LO) \cite{Ashcroft1976}. The optical response of the substrate is represented by a complex permittivity $\varepsilon (\tilde{\nu})$, whose real and imaginary parts are shown in the inset of Figure 2b as obtained from literature data \cite{Palik1997}. We point out that for plane wave excitation at normal incidence only the TO mode can directly couple to the incident light. Thus, the imaginary part of $\varepsilon$ peaks at the TO phonon frequency, giving the main contribution to the absorption in the SiO$_2$ layer. Furthermore, the real part of $\varepsilon$ is negative in the frequency range between the TO and the LO modes, the so-called  \textsl{reststrahlen} band, leading to a strong reflection at the air/SiO$_2$ interface.\\
Using $\varepsilon(\tilde{\nu})$ as input, we obtain the absorption from incident light passing through 300 nm of SiO$_2$, as shown in Figure 2b.

\begin{figure}[H]
\includegraphics[scale=1]{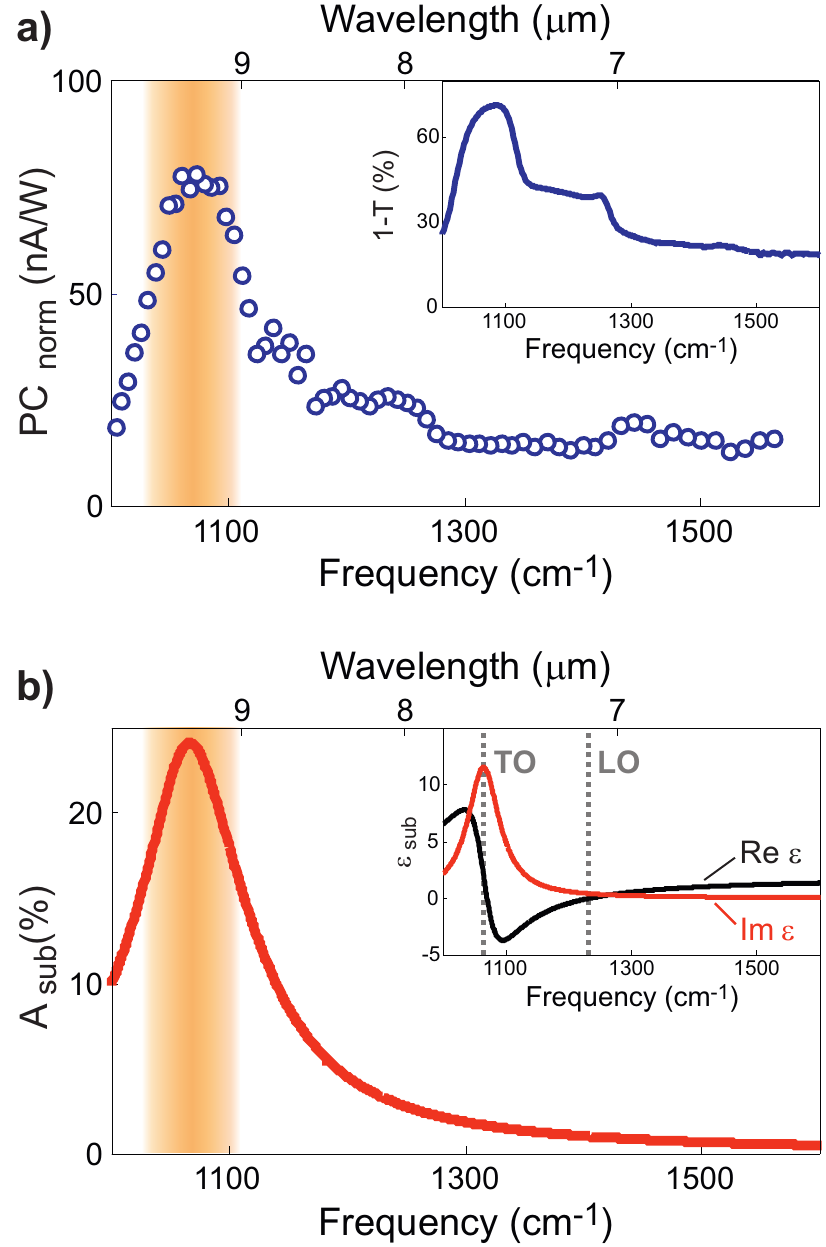}
\label{fig2}
\caption{(a) Spectrum of the normalized photocurrent, obtained by taking the maximum $PC_{\rm{norm}}$ from the spatial map, and for a gate voltage that is away from the charge-neutrality point ($V_{\rm{BG}}-V_{\rm{D}}$=70 V). Inset: extinction spectrum $1-T$ of the substrate, obtained by FTIR. (b) Calculated absorption in 300 nm SiO$_2$ taking into account the full layered SiO$_2$/Si/SiO$_2$ structure. Inset: real and imaginary part of the SiO$_2$ permittivity obtained from a fit of literature data \cite{Palik1997} to the expression: $\varepsilon(\tilde{\nu})=\varepsilon_{\infty}(1+\frac{{\tilde{\nu}_{LO}}^2-{\tilde{\nu}_{TO}}^2}{{\tilde{\nu}_{TO}}^2-{\tilde{\nu}}^2-i\tilde{\nu}\Gamma}$), with $\varepsilon_\infty=1.843$, $\tilde{\nu}_{LO}=1243.5$ \cm, $\tilde{\nu}_{TO}=1065.5$ \cm~ and $\Gamma=61.6$ \cm.}
\end{figure}
\noindent
We take into account the entire layered structure, since the penetration depth at $\tilde{\nu}_{\rm{e}}=$1080 \cm~ is $\sim$~300 nm, while for higher frequencies it is larger than the oxide thickness ($\sim$~10 \um~ for $\tilde{\nu}_{\rm{e}}=$1390 \cm, more details in the Supplementary Information). The calculated absorption peaks at the TO phonon and goes to zero for frequencies higher than the LO phonon. We thus observe a correlation between the substrate absorption and the measured photocurrent spectrum, as they both show a peak at the TO phonon band. The similarity is evident even if the frequency and width of the phonon modes in our samples differs slightly from literature values due to variations in growth conditions and thickness of the oxide.\\
We explain the relationship between the MIR light absorption in the SiO$_2$ and the photocurrent in the following way: after light absorption, heat is generated in the substrate, where it diffuses and equilibrates with the graphene. Consequently, when light is absorbed around the source electrode, an imbalance in the spatial distribution of the temperature of the graphene carriers is created, as the graphene close to the drain electrode ($\geq$100 \um~ away) is not heated. As a consequence of the carrier temperature difference in the source and drain regions, a thermovoltage is generated, governed by the Seebeck coefficient of graphene: $\Delta V= S (\theta_{\rm{s}}-\theta_{\rm{d}})$, where $S$ is the graphene Seebeck coefficient, and $\theta_{\rm{s}}$ ($\theta_{\rm{d}}$) is the temperature of the graphene charge carriers near the source (drain) contact. The large spatial extent of the photoresponse map as shown in Figure 1a can thus be related to the temperature distribution in the substrate.\\ 
We note that this non-local photo-thermoelectric mechanism is markedly different from the reported photo-thermoelectric response for VIS and NIR light near metallic contacts \cite{Freitag2013b} or interfaces, such as p-n junctions \cite{Gabor2011,Lemme2011a} or single-bilayer graphene \cite{Xu2010}. In those cases, the graphene carriers are directly excited by the laser and the temperature gradient is generated within or close to the laser spot. In our case, because the region with an elevated substrate temperature extends over a larger area than the spot size, it is possible to observe a photoresponse even when the laser spot is outside the graphene region.\\

In addition to this spatially extended photoresponse, we also observe a local photoresponse near the contacts, for which we show an example in Figure 1b. This local photoresponse is particularly clear for $\tilde{\nu}_{\rm{e}}>\tilde{\nu}_{\rm{LO}}$, where substrate absorption is nearly zero, but a significant photoresponse is still observed (Figure 2a). We attribute this local photoresponse to direct light absorption in the graphene, and verify the occurrence of graphene absorption by directly measuring the light transmission $T$ through the device. In Figures 3a,b we show $1-T$ as a function of backgate voltage for graphene upon 1080 \cm~ and 1390 \cm~ excitation, compared with the respective bare substrate extinction. 

\begin{figure}[H]
\includegraphics[scale=1]{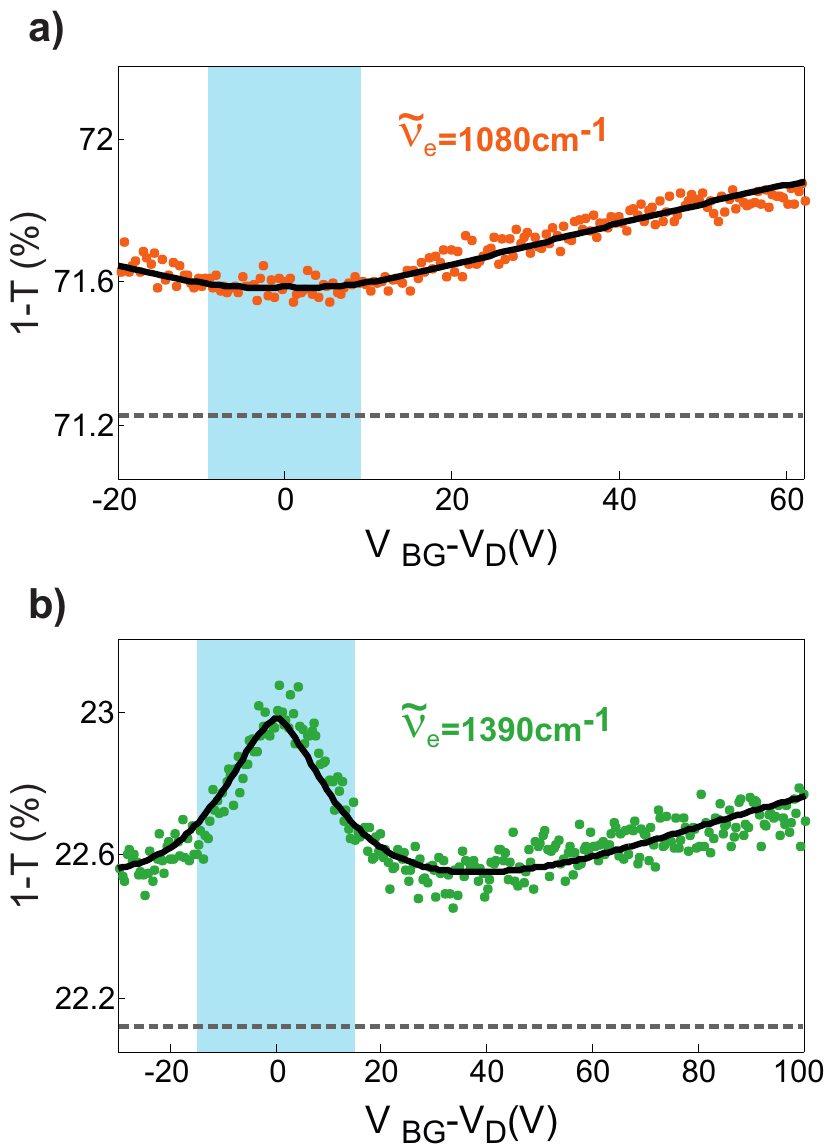}
\label{fig3}
\caption{Dependence of $1-T$ on gate voltage (where $T$ is the transmission) of the graphene+substrate system (symbols) or of the substrate alone (grey dashed lines), for $\tilde{\nu}=$1080 \cm~ (a), and for $\tilde{\nu}=$1390 \cm~ (b). The black solid curves are obtained from the model described in Methods. The light blue areas indicate the regions where interband electron-hole pair transitions are allowed.}
\end{figure}
\noindent
We compare the data with a simple model, where we calculate the graphene transmission starting from the optical conductivity of graphene at room temperature taking into account the spatially varying $E_F$ within our focus spot \cite{Martin2007a} (see details in Methods). We obtain good agreement with the use of one fitting parameter that represents the fraction of transmitted light through all the other layers of the substrate. The model includes both interband and intraband processes. The interband transitions are responsible for the well-known 2.3\% absorption for suspended graphene in the VIS and NIR \cite{Nair2008}, and occur for $E_0=\hbar\omega>2E_F$, while for $E_0$<2$E_F$ they are forbidden by Pauli blocking \cite{Mak2008a, Li2008}. Intraband scattering leads to Drude absorption, which at MIR frequencies is stronger for shorter carrier scattering time $\tau$ (notice that $\omega\tau > 1$) and for larger $E_F$. This mechanism is responsible for the absorption in the far-infrared (THz frequencies) \cite{Dawlaty2008, Yan2011} and also contributes to the MIR optical response \cite{Mak2008a, Dawlaty2008}. We note that absorption in the Pauli blocking region has been attributed also to electron-electron interaction \cite{Li2008}, whereas phonon-assisted absorption has been predicted for graphene on polar substrates \cite{Scharf2013}. \\
In our experiments the photon energy $E_0$ is in the 123-198  meV interval, and thus interband absorption is relevant only for a limited range of Fermi energies ($\sim$ 60-100 meV) around the charge-neutrality point, indicated in Figures 3a,b with light-blue areas. Outside the interband windows, we clearly observe the signatures of intraband absorption. Our absorption measurements, in combination with the photocurrent measurements, which reveal a response for a wide voltage range (also seen in Figure 5a, and discussed later), suggest that the photoresponse is also originating from direct light absorption in graphene.\\

To further investigate the contribution of absorption in graphene to photocurrent generation, we analyse the effect of the polarization of the incident light, comparing the cases where light polarization is parallel and perpendicular to the edge of the gold electrode. In Figure 4a we present spectra for a cross-shaped 4 terminal device, where we perform measurements for both polarizations without physically rotating the light polarization or the sample (see Supplementary Information). Strikingly, the spectrum of the absolute value of the photocurrent $|PC_{\rm{norm}}|$ for perpendicular polarization presents two peaks: one whose position corresponds to the peak of the photocurrent spectrum for parallel polarization, at the TO phonon frequency, and a second one at around 1170 \cm, approximately in the middle of the \textsl{reststrahlen} band.\\
Moreover, the spatial extent of the photocurrent signal is remarkably different for the two peak frequencies, as shown in the inset of Figure 4a for the frequency range $\tilde{\nu}_{\rm{TO}}\lesssim \tilde{\nu}_{\rm{e}} \lesssim \tilde{\nu}_{\rm{LO}}$. For perpendicular polarization, this spatial extent is comparable to the spot size, while for parallel polarization it is more than twice as large as the spot size. This suggests that the strong enhancement of the photoresponse for perpendicular polarization at $\tilde{\nu}_{\rm{e}} \approx 1170$ \cm~ is due to a local effect. We point out that in the case presented here the two peaks also have different signs (see raw data in the Supplementary Information). The reason why the photocurrent direction is opposite in the different spectral regions is not fully understood, but we speculate that this is due to changes in the local Seebeck coefficient for different spatial variations of the doping level. We observe in other samples that the presence of a sign change depends strongly on the graphene intrinsic doping and also on the position along the contact. We address this issue in details in the Supplementary Information.\\
A physical picture of our results emerges with the observation that the photoresponse peak at $\tilde{\nu}_{\rm{e}} \approx 1170$ \cm~  correlates with the energy of the surface optical phonon (SO) of the substrate. Analogous to surface plasmons in metals, surface phonons are evanescent waves originating from the ionic motion at the surface of polar materials; more precisely, they arise at the interface between two dielectric materials with permittivities of opposite signs, and their in-plane momenta are higher than the free-space wave-vector for light of the same frequency \cite{Kliewer1966}.\\

\begin{figure*}[t!!!!!]
\includegraphics[scale=1]{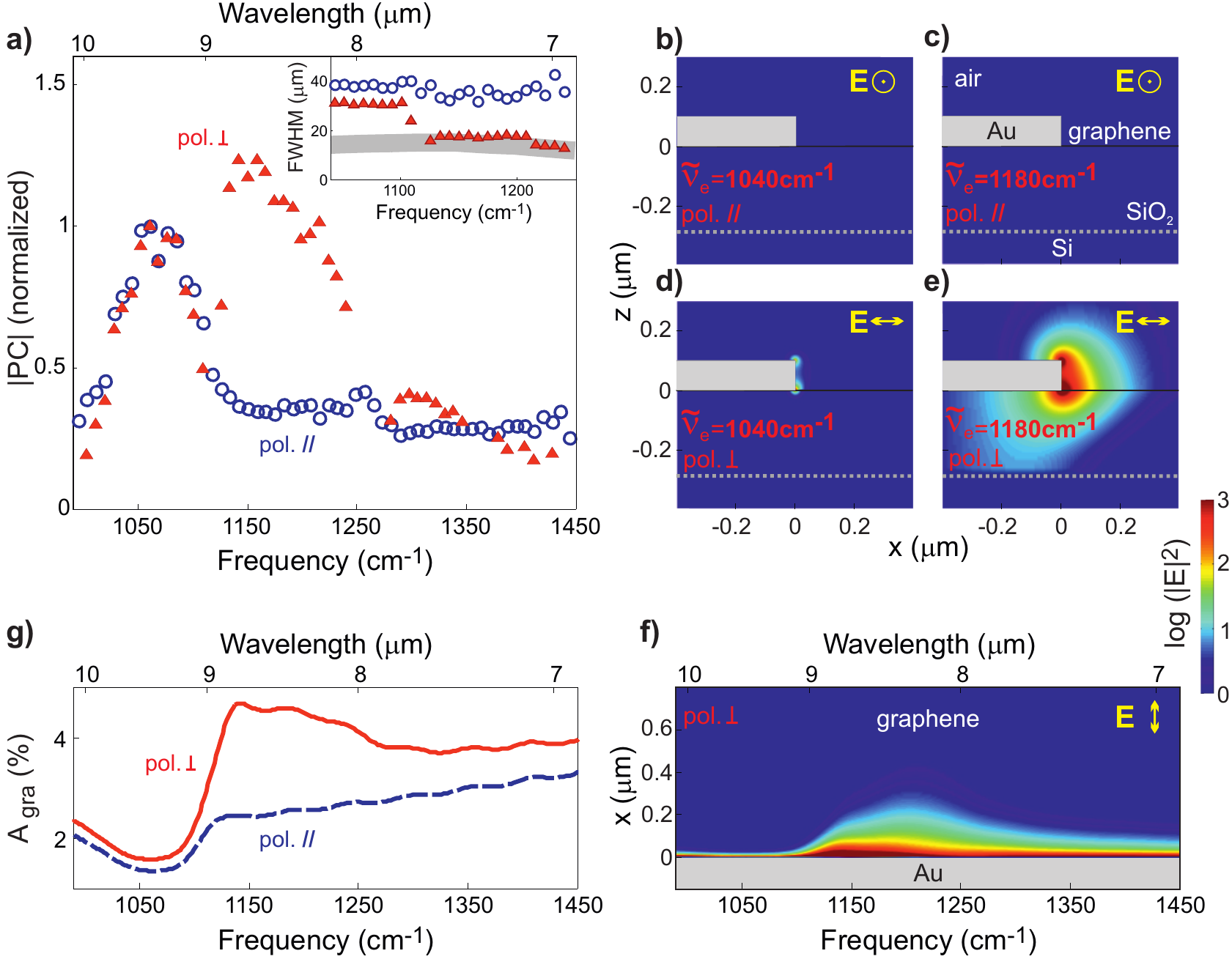}
\label{fig4}
\caption{(a) Photocurrent spectra for perpendicular (red triangles) and parallel (blue circles) polarization for a gate voltage that is away from the charge-neutrality point ($V_{\rm{BG}}-V_{\rm{D}}$=-90 V). Inset: corresponding fwhm of the photocurrent signal. The grey line represents the spot size width as retrieved from the transmission. (b),(c),(d),(e) Magnitude of the electric field from FDTD simulations. Side view of the device, the dimensions used are  100 nm thick Au, 300 nm thick SiO$_2$, and semi-infinite Si. The frequency dependent permittivities of the involved materials are taken from literature \cite{Palik1997}, and in the case of SiO$_2$ it includes the phonon modes. (b): with $\tilde{\nu}_{\rm{e}}$=1040 \cm~ with parallel polarization; (c): with $\tilde{\nu}_{\rm{e}}$=1180 \cm~ with parallel polarization; (d): with $\tilde{\nu}_{\rm{e}}$=1040 \cm~ with perpendicular polarization; (e): with $\tilde{\nu}_{\rm{e}}$=1080 \cm~ with perpendicular polarization. (f) Magnitude of the electric field in the graphene plane from FDTD simulations under illumination with light perpendicular to the contact edge as a function of frequency and distance from the metal edge. Top view of the device. (g) Comparison between the graphene absorption from FDTD simulations for light perpendicular and parallel to the contact integrated over 10 \um~ from the edge, for $E_F$=0.1 eV.}
\end{figure*}

\newpage
\noindent
In a simple electrostatic picture, when an electric field impinges perpendicularly on a metal edge, there is an accumulation of charges at the edge, resulting in a strong field perpendicular to the conductor surface. Hence, the near field at the top and bottom corners of the metal contact carries a distribution of large in-plane momenta. The presence of large in-plane momenta thus enables the excitation of the surface phonons when the frequency of the incoming light matches the surface phonon resonance frequency $\tilde{\nu}_{\rm{SO}}$, as shown in near-field measurements \cite{Hillenbrand2002}. In turn, this field enhancement leads to an enhancement of the absorption of the incoming light in the graphene sheet near the interface, mediated by the excitation of substrate surface phonons.\\
In order to better understand the implications of this effect, we perform simulations of the electric field at the interface between graphene and gold on SiO$_2$ by numerically solving Maxwell's equations with a finite-difference time-domain method in two dimensions (FDTD, using the software Lumerical). In Figures 4b,c,d,e we show a side view of the magnitude of the electric field close to a contact for four different situations: in Figures 4b,c under illumination with light with parallel polarization, with frequency far from (1040 \cm, 4b) and close (1180 \cm, 4c) to $\tilde{\nu}_{\rm{SO}}$; in Figures 4d,e under illumination with light with the same two frequencies but polarization perpendicular to the contact edge. We observe very distinct features in both field magnitude and spatial profile. In the cases where the polarization is parallel to the gold edge, there is no electric field enhancement, as expected, since the only effect of the electric field in this case would be moving charges in the parallel direction. Instead, when the light polarization is perpendicular to the edge, electric field localization occurs at the corners of the contact. When the light frequency is close to $\tilde{\nu}_{\rm{SO}}$, the excitation of SO phonons results in a strong field enhancement.\\
The extent of the field enhancement in the graphene plane (for light polarization perpendicular to the contact edge) is presented in Figure 4f for the range of frequencies spanned in the photocurrent spectrum. In Figure 4g we show the graphene absorption at $E_F$=0.1 eV integrated over 10 \um~ from the contact, a distance comparable to the beam spot size, for light polarizations parallel and perpendicular to the contact edge. We observe two interesting features: first, the graphene absorption is strongly enhanced in the frequency range where increased photocurrent is obtained for perpendicular  polarization. This observation further supports a scenario in which the graphene absorption in proximity of the graphene/gold interface is strongly enhanced due to the electric field localization produced by the excitation of the SO phonon modes, thus resulting in an increase of the generated photocurrent. We remark that for both polarizations there is a dip in the graphene absorption at the TO phonon resonance. This can be understood in terms of graphene absorption on the substrate, which is proportional to $\vert t \vert^2 $, where $t$ is the Fresnel transmission coefficient. Because $\vert t \vert^2 \propto 1/\mbox{Im}(\varepsilon)$, the absorption has a minimum at the TO phonon resonance. This observation also supports the assignment of the photocurrent peak observed at this frequency to substrate-absorption-mediated photocurrent.\\

Finally, we study the effect of the Fermi energy on the photoresponse for the different regimes and explain the results within a simple thermoelectric model. The backgate dependences of the normalized photoresponse for 1080 \cm~ and 1390 \cm~ illumination with parallel polarization are shown in Figure 5a. We observe that the photocurrent generated by the 1080 \cm~ light (i.e. on resonance with the TO phonon of the substrate) is significantly stronger over the entire applied voltage range and relatively symmetrical with respect to $V_{\rm{BG}}=V_{\rm{D}}$. The photocurrent at frequencies outside the TO phonon resonance exhibits asymmetrical behaviour with respect to $V_{\rm{D}} $: it is strongly negative for $V_{\rm{BG}}-V_{\rm{D}}<0$, and only weakly positive for higher voltages. We give a qualitative explanation for these different observed behaviors using a single framework based on the thermoelectric effect, where the difference arises due to the spatial extent of the temperature distribution, which is much larger for excitation at 1080 \cm, compared to excitation at 1390 \cm. We compute the thermoelectric photocurrent originating from the two different temperature distributions $\theta(x)$ at the two different excitation frequencies, by representing the device by a simple Seebeck coefficient profile, as illustrated in the inset of Figure 5b. We consider three regions around the gold/graphene contact: the gold, with fixed Seebeck coefficient $S_{\rm{Au}}$; the graphene, whose Fermi level is pinned by the gold, with fixed Seebeck coefficient $S_{\rm{g/Au}}$; and the graphene with gate-tunable Seebeck coefficient $S_{\rm{g}}(V_{\rm{BG}})$ obtained via the Mott formula \cite{Ashcroft1976, Zuev2009a, Wei2009}. \\
\begin{figure}[H]
\includegraphics[scale=1]{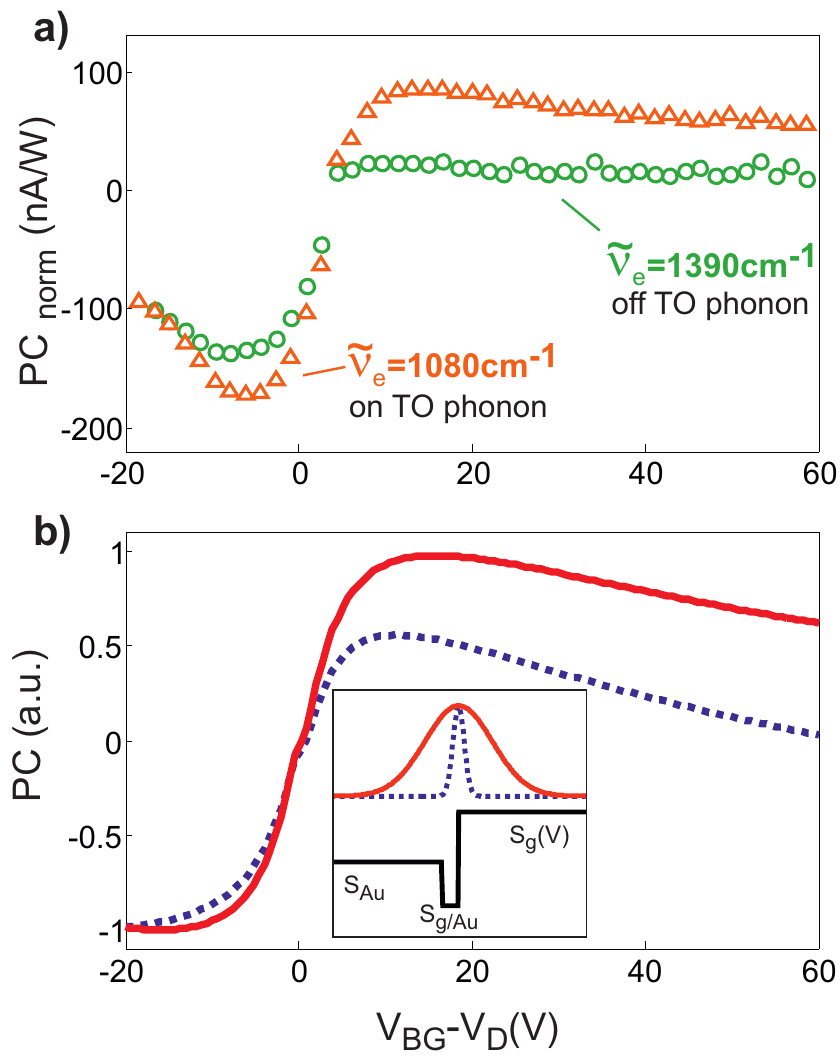}
\label{fig5}
\caption{(a) Photocurrent as a function of backgate voltage under illumination with $\tilde{\nu}_{\rm{e}}$=1080 \cm~ light (orange triangles), and $\tilde{\nu}_{\rm{e}}$=1390 \cm~ light (green circles). (b) Calculated photocurrent (normalized to the absolute value of the minimum) obtained from the integration of $PC$ from the equation \ref{pc}, using the $\theta(x)$ distributions shown in the inset: red solid curve for the large width one, blue dotted curve for the one comparable to the spot size. The parameters used to estimate the Seebeck coefficient and the resistance are: mobility $\mu$=2000 $\rm{cm}^2$/Vs, residual charge density at the Dirac point $\rm{n_0}$=2$\cdot \rm{10^{11}~cm^{-2}}$, contact resistance $\rm{R_c}$=1000 $\Omega$,  $S_{\rm{g/Au}}=S_{\rm{g}}(V_{\rm{BG}}=1~\rm{V})$}
\end{figure}

\noindent
The photocurrent is thus proportional to the spatial integral of the product of the Seebeck coefficient and the temperature gradient,
\begin{equation}
\label{pc}
PC\propto \frac{1}{R}\int S(x)\nabla \theta(x) dx 
\end{equation} 
where $R$ is the device resistance. We describe these two mechanisms by means of Gaussian temperature distributions with different widths. We assume that the absorption in the graphene gives rise to a temperature profile whose width is comparable to the laser spot size and to the width of the gold-induced doping region with $S_{\rm{g/Au}}$. In contrast, the temperature distribution arising from absorption in the substrate can be five times larger, following the observations in Figure 1a. As we can see in Figure 5b, where we show the results of the simulations for the set of parameters that best describes our data (see caption), this very simple model captures the main features observed in Figure 5a: the photocurrent resulting from a temperature distribution with a large spatial extent is symmetric with respect to $V_{\rm{D}}$ (i.e. the negative and positive responses are comparable in magnitude). Indeed, as mentioned above, this is due to the fact that in this case the main contribution to the backgate dependence comes from the graphene region in between the contacts, hence it is directly proportional to $S_{\rm{g}}(V_{\rm{BG}})$. Instead, when the temperature distribution width is comparable to or smaller than the the gold-induced doping region, the photocurrent is mainly created at the graphene-graphene junction near the contact, and thus defined by the $S_{\rm{g/Au}}$ and $S_{\rm{g}}(V_{\rm{BG}})$, which results in a strongly asymmetrical signal. Such photo-thermoelectric response has been reported before for visible light, impinging on graphene pn-junctions \cite{Gabor2011,Lemme2011a}.\\

In conclusion, we infer the presence of different mechanisms of photocurrent generation in graphene on a polar substrate under excitation with MIR light. One of them is mediated by substrate absorption: the signal peaks at a frequency corresponding to the TO phonon resonance, and shows a spatial extent larger than the beam spot size. The backgate voltage dependence of the photoresponse is well described with a simple model based upon the photo-thermoelectric effect. The other mechanism is due to hot carrier generation via absorption in graphene. In this case the photocurrent is localized to the contact/graphene interface, where strong field enhancement via excitation of the substrate SO phonons can greatly increase the response. Understanding the role of these photocurrent generation mechanisms paves the way to the possibility of tailoring the magnitude and spatial extent of the graphene photo-response, for example by engineering the substrate permittivity or by enhancing the electric field localization. Our results open new routes for using graphene in compact and low-cost mid-infrared sensors and imaging systems operating at room-temperature.

\section*{Methods}
\textbf{Experimental}\\
The samples consist of CVD graphene deposited onto a SiO$_2$ (300 nm)/Si or SiO$_2$ (300 nm)/Si/SiO$_2$ (300 nm) substrate and etched into rectangles with $\geq$ 100 \um~ between the Ti (3 nm)/Au (100 nm) electrodes. In order to optimize the transmission of light in the $\sim$6-10 \um~  wavelength range, we use double-side polished substrates to reduce scattering from the back surface, and relatively low-doped Si (sheet resistance 5-10 $\Omega/\rm{cm}$) to reduce the effect of the (Drude) absorption of the charge carriers in the Si, while still being able to efficiently gate the device. Device 1 is used in Figure 1, device 2 is used in Figure 2, 3 and 5 and device 3 is used in Figure 4.\\
The scanning photocurrent images are collected by focusing the laser beam with ZnSe lenses, and moving the sample mounted on a motorized stage. We modulate the laser light with a chopper at 423 Hz frequency. The photocurrent is amplified by a Femto DLPCA-200 preamplifier and the lock-in signal is obtained by a Stanford Research Systems SR830 DSP. A lock-in amplifier (Femto LIA-MV-150) is used for the transmission measurements. The experiments are carried out at room temperature in $\rm{N_2}$ atmosphere, in order to minimize the effects of the absorption by air in our wavelength range. The output power of the laser varies for each wavelength and is in the mW range. The linear dependence of the photocurrent on incident power is tested (see Supplementary Information).\\
\textbf{Model for the transmission curves}\\
In the limit of graphene on an infinite SiO$_2$ substrate, the transmission defined via the Poynting vector after having passed a distance $W$ is given by $T= n \vert t \vert ^2  e^{-\frac{4\pi \kappa W}{\lambda}}$, where $t = 2/(1 + \sqrt\varepsilon +\frac{\sigma}{\epsilon_0 c})$, $\sqrt\varepsilon = n + i\kappa$ and $\lambda$ is the light wavelength. This is an approximation to the full system, where the effect of multiple reflections in the SiO$_2$ and Si layers is neglected. The study of the graphene absorption considering the full multilayer system is presented in the Supplementary Information. However, we are interested in a simple model that can describe our data without relying heavily on the material parameters taken from literature which might be sample dependent. This procedure is further justified by the fact that we concentrate only on the term $\vert t \vert ^2$ which bears the graphene dependence on $E_F$. We then compute $\vert t \vert ^2$ using $\sigma(\tilde{\nu},E_F)=\sigma_{intra}(\tilde{\nu},E_F)+\sigma_{inter}(\tilde{\nu},E_F)$, where $\sigma_{intra}(\tilde{\nu},E_F)$ and $\sigma_{inter}(\tilde{\nu},E_F)$ are obtained from the literature \cite{Falkovsky2007, Falkovsky2008, Wunsch2006, Hwang2007a}. We take the scattering time as $\tau~$= 20 fs, which is a realistic scattering time considering that our samples mobility is of the order of 1000 $\rm{cm^2/Vs}$. We probe an inhomogeneous sample with a $\sim$10 \um~ light spot: we include the effect of Fermi level variation within the spot-size by a conductivity averaged over a distribution of Fermi levels with a standard deviation of 0.12 eV. To explain our data we use :
\[ T_{total}(\tilde{\nu},E_F)=\vert t \vert ^2 \cdot T_{sub} \] and then obtain $T_{sub}$ by imposing that the value at $\rm{V_D}$ be the same as the smoothed data.

\section*{Associated content}
\textbf{Supplementary Information}\\
Dependence of the photocurrent on the excitation power, details on the perpendicular polarization results, model for the photocurrent backgate dependence and calculation of graphene absorption on a polar substrate. This material is available free of charge via the Internet at http://pubs.acs.org.

\section*{Acknowledgements}
We are grateful for discussions with Pablo Jarillo-Herrero, Marko Spasenovi\'c, Qiong Ma, Romain Parret and Guillermo G\'omez-Santos. It is a pleasure to thank Davide Janner for precious support in the clean room, and Lorenzo Giannini for the artistic sketch. T.S. acknowledges support by the project FIS2013-44098. F.K. acknowledges support by the Fundacio Cellex Barcelona, the ERC Career integration grant 294056 (GRANOP), the ERC starting grant 307806 (CarbonLight) and the Accion Integrada-MINECO grant (AAII- PRI-AIBSE-2011-1298). J.G.A. and F.K. acknowledge support by the E.C. under Graphene Flagship (contract no. CNECT-ICT-604391), and the European project GRASP.

\providecommand{\latin}[1]{#1}
\providecommand*\mcitethebibliography{\thebibliography}
\csname @ifundefined\endcsname{endmcitethebibliography}
  {\let\endmcitethebibliography\endthebibliography}{}


\begin{mcitethebibliography}{45}
\providecommand*\natexlab[1]{#1}
\providecommand*\mciteSetBstSublistMode[1]{}
\providecommand*\mciteSetBstMaxWidthForm[2]{}
\providecommand*\mciteBstWouldAddEndPuncttrue
  {\def\EndOfBibitem{\unskip.}}
\providecommand*\mciteBstWouldAddEndPunctfalse
  {\let\EndOfBibitem\relax}
\providecommand*\mciteSetBstMidEndSepPunct[3]{}
\providecommand*\mciteSetBstSublistLabelBeginEnd[3]{}
\providecommand*\EndOfBibitem{}
\mciteSetBstSublistMode{f}
\mciteSetBstMaxWidthForm{subitem}{(\alph{mcitesubitemcount})}
\mciteSetBstSublistLabelBeginEnd
  {\mcitemaxwidthsubitemform\space}
  {\relax}
  {\relax}

\bibitem[Bonaccorso \latin{et~al.}(2010)Bonaccorso, Sun, Hasan, and
  Ferrari]{Bonaccorso2010a}
Bonaccorso,~F.; Sun,~Z.; Hasan,~T.; Ferrari,~A.~C. \emph{Nature Photonics}
  \textbf{2010}, \emph{4}, 611--622\relax
\mciteBstWouldAddEndPuncttrue
\mciteSetBstMidEndSepPunct{\mcitedefaultmidpunct}
{\mcitedefaultendpunct}{\mcitedefaultseppunct}\relax
\EndOfBibitem
\bibitem[Ferrari \latin{et~al.}(2015)Ferrari, Bonaccorso, Fal{'}ko, Novoselov,
  Roche, Boggild, Borini, Koppens, Palermo, Pugno, Garrido, Sordan, Bianco,
  Ballerini, Prato, Lidorikis, Kivioja, Marinelli, Ryhanen, Morpurgo, Coleman,
  Nicolosi, Colombo, Fert, Garcia-Hernandez, Bachtold, Schneider, Guinea,
  Dekker, Barbone, Sun, Galiotis, Grigorenko, Konstantatos, Kis, Katsnelson,
  Vandersypen, Loiseau, Morandi, Neumaier, Treossi, Pellegrini, Polini,
  Tredicucci, Williams, Hee~Hong, Ahn, Min~Kim, Zirath, van Wees, van~der Zant,
  Occhipinti, Di~Matteo, Kinloch, Seyller, Quesnel, Feng, Teo, Rupesinghe,
  Hakonen, Neil, Tannock, Lofwander, and Kinaret]{Ferrari2014}
Ferrari,~A.~C. \latin{et~al.}  \emph{Nanoscale} \textbf{2015}, \emph{7},
  4598--4810\relax
\mciteBstWouldAddEndPuncttrue
\mciteSetBstMidEndSepPunct{\mcitedefaultmidpunct}
{\mcitedefaultendpunct}{\mcitedefaultseppunct}\relax
\EndOfBibitem
\bibitem[Grigorenko \latin{et~al.}(2012)Grigorenko, Polini, and
  Novoselov]{Grigorenko2012a}
Grigorenko,~A.~N.; Polini,~M.; Novoselov,~K.~S. \emph{Nature Photonics}
  \textbf{2012}, \emph{6}, 749--758\relax
\mciteBstWouldAddEndPuncttrue
\mciteSetBstMidEndSepPunct{\mcitedefaultmidpunct}
{\mcitedefaultendpunct}{\mcitedefaultseppunct}\relax
\EndOfBibitem
\bibitem[{Garc\'{\i}a de Abajo}(2014)]{GarciadeAbajo2014a}
{Garc\'{\i}a de Abajo},~F.~J. \emph{ACS Photonics} \textbf{2014}, \emph{1},
  135--152\relax
\mciteBstWouldAddEndPuncttrue
\mciteSetBstMidEndSepPunct{\mcitedefaultmidpunct}
{\mcitedefaultendpunct}{\mcitedefaultseppunct}\relax
\EndOfBibitem
\bibitem[Low and Avouris(2014)Low, and Avouris]{Low2014c}
Low,~T.; Avouris,~P. \emph{ACS Nano} \textbf{2014}, \emph{8}, 1086--101\relax
\mciteBstWouldAddEndPuncttrue
\mciteSetBstMidEndSepPunct{\mcitedefaultmidpunct}
{\mcitedefaultendpunct}{\mcitedefaultseppunct}\relax
\EndOfBibitem
\bibitem[Koppens \latin{et~al.}(2014)Koppens, Mueller, Avouris, Ferrari,
  Vitiello, and Polini]{Koppens2014}
Koppens,~F. H.~L.; Mueller,~T.; Avouris,~P.; Ferrari,~A.~C.; Vitiello,~M.~S.;
  Polini,~M. \emph{Nature Nanotechnology} \textbf{2014}, \emph{9},
  780--793\relax
\mciteBstWouldAddEndPuncttrue
\mciteSetBstMidEndSepPunct{\mcitedefaultmidpunct}
{\mcitedefaultendpunct}{\mcitedefaultseppunct}\relax
\EndOfBibitem
\bibitem[Chen \latin{et~al.}(2008)Chen, Jang, Xiao, Ishigami, and
  Fuhrer]{Chen2008}
Chen,~J.-H.; Jang,~C.; Xiao,~S.; Ishigami,~M.; Fuhrer,~M.~S. \emph{Nature
  Nanotechnology} \textbf{2008}, \emph{3}, 206--209\relax
\mciteBstWouldAddEndPuncttrue
\mciteSetBstMidEndSepPunct{\mcitedefaultmidpunct}
{\mcitedefaultendpunct}{\mcitedefaultseppunct}\relax
\EndOfBibitem
\bibitem[Meric \latin{et~al.}(2008)Meric, Han, Young, Ozyilmaz, Kim, and
  Shepard]{Meric2008}
Meric,~I.; Han,~M.~Y.; Young,~A.~F.; Ozyilmaz,~B.; Kim,~P.; Shepard,~K.~L.
  \emph{Nature Nanotechnology} \textbf{2008}, \emph{3}, 654--9\relax
\mciteBstWouldAddEndPuncttrue
\mciteSetBstMidEndSepPunct{\mcitedefaultmidpunct}
{\mcitedefaultendpunct}{\mcitedefaultseppunct}\relax
\EndOfBibitem
\bibitem[Hwang and {Das Sarma}(2013)Hwang, and {Das Sarma}]{Hwang2013}
Hwang,~E.; {Das Sarma},~S. \emph{Physical Review B} \textbf{2013}, \emph{87},
  115432\relax
\mciteBstWouldAddEndPuncttrue
\mciteSetBstMidEndSepPunct{\mcitedefaultmidpunct}
{\mcitedefaultendpunct}{\mcitedefaultseppunct}\relax
\EndOfBibitem
\bibitem[Wang and Mahan(1972)Wang, and Mahan]{Wang1972}
Wang,~S.; Mahan,~G. \emph{Physical Review B} \textbf{1972}, \emph{6},
  4517--4524\relax
\mciteBstWouldAddEndPuncttrue
\mciteSetBstMidEndSepPunct{\mcitedefaultmidpunct}
{\mcitedefaultendpunct}{\mcitedefaultseppunct}\relax
\EndOfBibitem
\bibitem[Fratini and Guinea(2008)Fratini, and Guinea]{Fratini2008}
Fratini,~S.; Guinea,~F. \emph{Physical Review B} \textbf{2008}, \emph{77},
  195415\relax
\mciteBstWouldAddEndPuncttrue
\mciteSetBstMidEndSepPunct{\mcitedefaultmidpunct}
{\mcitedefaultendpunct}{\mcitedefaultseppunct}\relax
\EndOfBibitem
\bibitem[Freitag \latin{et~al.}(2013)Freitag, Low, and Avouris]{Freitag2013b}
Freitag,~M.; Low,~T.; Avouris,~P. \emph{Nano Letters} \textbf{2013}, \emph{13},
  1644--8\relax
\mciteBstWouldAddEndPuncttrue
\mciteSetBstMidEndSepPunct{\mcitedefaultmidpunct}
{\mcitedefaultendpunct}{\mcitedefaultseppunct}\relax
\EndOfBibitem
\bibitem[Low \latin{et~al.}(2012)Low, Perebeinos, Kim, Freitag, and
  Avouris]{Low2012}
Low,~T.; Perebeinos,~V.; Kim,~R.; Freitag,~M.; Avouris,~P. \emph{Physical
  Review B} \textbf{2012}, \emph{86}, 045413\relax
\mciteBstWouldAddEndPuncttrue
\mciteSetBstMidEndSepPunct{\mcitedefaultmidpunct}
{\mcitedefaultendpunct}{\mcitedefaultseppunct}\relax
\EndOfBibitem
\bibitem[Yan \latin{et~al.}(2012)Yan, Kim, Elle, Sushkov, Jenkins, Milchberg,
  Fuhrer, and Drew]{Yan2012a}
Yan,~J.; Kim,~M.-H.; Elle,~J.~A.; Sushkov,~A.~B.; Jenkins,~G.~S.;
  Milchberg,~H.~M.; Fuhrer,~M.~S.; Drew,~H.~D. \emph{Nature Nanotechnology}
  \textbf{2012}, \emph{7}, 472--8\relax
\mciteBstWouldAddEndPuncttrue
\mciteSetBstMidEndSepPunct{\mcitedefaultmidpunct}
{\mcitedefaultendpunct}{\mcitedefaultseppunct}\relax
\EndOfBibitem
\bibitem[Freitag \latin{et~al.}(2013)Freitag, Low, Zhu, Yan, Xia, and
  Avouris]{Freitag2013}
Freitag,~M.; Low,~T.; Zhu,~W.; Yan,~H.; Xia,~F.; Avouris,~P. \emph{Nature
  Communications} \textbf{2013}, \emph{4}, 1951\relax
\mciteBstWouldAddEndPuncttrue
\mciteSetBstMidEndSepPunct{\mcitedefaultmidpunct}
{\mcitedefaultendpunct}{\mcitedefaultseppunct}\relax
\EndOfBibitem
\bibitem[Yao \latin{et~al.}(2014)Yao, Shankar, Rauter, Song, Kong, Loncar, and
  Capasso]{Yao2014}
Yao,~Y.; Shankar,~R.; Rauter,~P.; Song,~Y.; Kong,~J.; Loncar,~M.; Capasso,~F.
  \emph{Nano Letters} \textbf{2014}, \emph{14}, 3749--54\relax
\mciteBstWouldAddEndPuncttrue
\mciteSetBstMidEndSepPunct{\mcitedefaultmidpunct}
{\mcitedefaultendpunct}{\mcitedefaultseppunct}\relax
\EndOfBibitem
\bibitem[Herring \latin{et~al.}(2014)Herring, Hsu, Gabor, Shin, Kong, Palacios,
  and Jarillo-Herrero]{Herring2014}
Herring,~P.~K.; Hsu,~A.~L.; Gabor,~N.~M.; Shin,~Y.~C.; Kong,~J.; Palacios,~T.;
  Jarillo-Herrero,~P. \emph{Nano Letters} \textbf{2014}, \emph{14},
  901--7\relax
\mciteBstWouldAddEndPuncttrue
\mciteSetBstMidEndSepPunct{\mcitedefaultmidpunct}
{\mcitedefaultendpunct}{\mcitedefaultseppunct}\relax
\EndOfBibitem
\bibitem[Freitag \latin{et~al.}(2014)Freitag, Low, Martin-Moreno, Zhu, Guinea,
  and Avouris]{Freitag2014}
Freitag,~M.; Low,~T.; Martin-Moreno,~L.; Zhu,~W.; Guinea,~F.; Avouris,~P.
  \emph{ACS nano} \textbf{2014}, \emph{8}, 8350--6\relax
\mciteBstWouldAddEndPuncttrue
\mciteSetBstMidEndSepPunct{\mcitedefaultmidpunct}
{\mcitedefaultendpunct}{\mcitedefaultseppunct}\relax
\EndOfBibitem
\bibitem[Jiang \latin{et~al.}(2011)Jiang, Shalygin, Panevin, Danilov, Glazov,
  Yakimova, Lara-Avila, Kubatkin, and Ganichev]{Jiang2011b}
Jiang,~C.; Shalygin,~V.~A.; Panevin,~V.~Y.; Danilov,~S.~N.; Glazov,~M.~M.;
  Yakimova,~R.; Lara-Avila,~S.; Kubatkin,~S.; Ganichev,~S.~D. \emph{Physical
  Review B} \textbf{2011}, \emph{84}, 125429\relax
\mciteBstWouldAddEndPuncttrue
\mciteSetBstMidEndSepPunct{\mcitedefaultmidpunct}
{\mcitedefaultendpunct}{\mcitedefaultseppunct}\relax
\EndOfBibitem
\bibitem[Olbrich \latin{et~al.}(2013)Olbrich, Drexler, Golub, Danilov,
  Shalygin, Yakimova, Lara-Avila, Kubatkin, Redlich, Huber, and
  Ganichev]{Olbrich2013}
Olbrich,~P.; Drexler,~C.; Golub,~L.~E.; Danilov,~S.~N.; Shalygin,~V.~a.;
  Yakimova,~R.; Lara-Avila,~S.; Kubatkin,~S.; Redlich,~B.; Huber,~R.;
  Ganichev,~S.~D. \emph{Physical Review B} \textbf{2013}, \emph{88},
  245425\relax
\mciteBstWouldAddEndPuncttrue
\mciteSetBstMidEndSepPunct{\mcitedefaultmidpunct}
{\mcitedefaultendpunct}{\mcitedefaultseppunct}\relax
\EndOfBibitem
\bibitem[Lee \latin{et~al.}(2008)Lee, Balasubramanian, Weitz, Burghard, and
  Kern]{Lee2008}
Lee,~E. J.~H.; Balasubramanian,~K.; Weitz,~R.~T.; Burghard,~M.; Kern,~K.
  \emph{Nature Nanotechnology} \textbf{2008}, \emph{3}, 486--90\relax
\mciteBstWouldAddEndPuncttrue
\mciteSetBstMidEndSepPunct{\mcitedefaultmidpunct}
{\mcitedefaultendpunct}{\mcitedefaultseppunct}\relax
\EndOfBibitem
\bibitem[Park \latin{et~al.}(2009)Park, Ahn, and Ruiz-Vargas]{Park2009a}
Park,~J.; Ahn,~Y.~H.; Ruiz-Vargas,~C. \emph{Nano Letters} \textbf{2009},
  \emph{9}, 1742--1746\relax
\mciteBstWouldAddEndPuncttrue
\mciteSetBstMidEndSepPunct{\mcitedefaultmidpunct}
{\mcitedefaultendpunct}{\mcitedefaultseppunct}\relax
\EndOfBibitem
\bibitem[Xia \latin{et~al.}(2009)Xia, Mueller, Golizadeh-Mojarad, Freitag, Lin,
  Tsang, Perebeinos, and Avouris]{Xia2009}
Xia,~F.; Mueller,~T.; Golizadeh-Mojarad,~R.; Freitag,~M.; Lin,~Y.-m.;
  Tsang,~J.; Perebeinos,~V.; Avouris,~P. \emph{Nano Letters} \textbf{2009},
  \emph{9}, 1039--44\relax
\mciteBstWouldAddEndPuncttrue
\mciteSetBstMidEndSepPunct{\mcitedefaultmidpunct}
{\mcitedefaultendpunct}{\mcitedefaultseppunct}\relax
\EndOfBibitem
\bibitem[Mueller \latin{et~al.}(2009)Mueller, Xia, Freitag, Tsang, and
  Avouris]{Mueller2009}
Mueller,~T.; Xia,~F.; Freitag,~M.; Tsang,~J.; Avouris,~P. \emph{Physical Review
  B} \textbf{2009}, \emph{79}, 245430\relax
\mciteBstWouldAddEndPuncttrue
\mciteSetBstMidEndSepPunct{\mcitedefaultmidpunct}
{\mcitedefaultendpunct}{\mcitedefaultseppunct}\relax
\EndOfBibitem
\bibitem[Gabor \latin{et~al.}(2011)Gabor, Song, Ma, Nair, Taychatanapat,
  Watanabe, Taniguchi, Levitov, and Jarillo-Herrero]{Gabor2011}
Gabor,~N.~M.; Song,~J. C.~W.; Ma,~Q.; Nair,~N.~L.; Taychatanapat,~T.;
  Watanabe,~K.; Taniguchi,~T.; Levitov,~L.~S.; Jarillo-Herrero,~P.
  \emph{Science} \textbf{2011}, \emph{334}, 648--52\relax
\mciteBstWouldAddEndPuncttrue
\mciteSetBstMidEndSepPunct{\mcitedefaultmidpunct}
{\mcitedefaultendpunct}{\mcitedefaultseppunct}\relax
\EndOfBibitem
\bibitem[Ashcroft and Mermin(1976)Ashcroft, and Mermin]{Ashcroft1976}
Ashcroft,~N.~W.; Mermin,~N.~D. \emph{{Solid State Physics}}; Harcourt College
  Publishers, 1976\relax
\mciteBstWouldAddEndPuncttrue
\mciteSetBstMidEndSepPunct{\mcitedefaultmidpunct}
{\mcitedefaultendpunct}{\mcitedefaultseppunct}\relax
\EndOfBibitem
\bibitem[Palik(1997)]{Palik1997}
Palik,~E.~D. \emph{{Handbook of Optical Constants of Solids}}; Elsevier: New
  York, 1997\relax
\mciteBstWouldAddEndPuncttrue
\mciteSetBstMidEndSepPunct{\mcitedefaultmidpunct}
{\mcitedefaultendpunct}{\mcitedefaultseppunct}\relax
\EndOfBibitem
\bibitem[Lemme \latin{et~al.}(2011)Lemme, Koppens, Falk, Rudner, Park, Levitov,
  and Marcus]{Lemme2011a}
Lemme,~M.~C.; Koppens,~F. H.~L.; Falk,~A.~L.; Rudner,~M.~S.; Park,~H.;
  Levitov,~L.~S.; Marcus,~C.~M. \emph{Nano Letters} \textbf{2011}, \emph{11},
  4134--7\relax
\mciteBstWouldAddEndPuncttrue
\mciteSetBstMidEndSepPunct{\mcitedefaultmidpunct}
{\mcitedefaultendpunct}{\mcitedefaultseppunct}\relax
\EndOfBibitem
\bibitem[Xu \latin{et~al.}(2010)Xu, Gabor, Alden, van~der Zande, and
  McEuen]{Xu2010}
Xu,~X.; Gabor,~N.~M.; Alden,~J.~S.; van~der Zande,~A.~M.; McEuen,~P.~L.
  \emph{Nano Letters} \textbf{2010}, \emph{10}, 562--566\relax
\mciteBstWouldAddEndPuncttrue
\mciteSetBstMidEndSepPunct{\mcitedefaultmidpunct}
{\mcitedefaultendpunct}{\mcitedefaultseppunct}\relax
\EndOfBibitem
\bibitem[Martin \latin{et~al.}(2007)Martin, Akerman, Ulbricht, Lohmann, Smet,
  von Klitzing, and Yacoby]{Martin2007a}
Martin,~J.; Akerman,~N.; Ulbricht,~G.; Lohmann,~T.; Smet,~J.~H.; von
  Klitzing,~K.; Yacoby,~A. \emph{Nature Physics} \textbf{2007}, \emph{4},
  144--148\relax
\mciteBstWouldAddEndPuncttrue
\mciteSetBstMidEndSepPunct{\mcitedefaultmidpunct}
{\mcitedefaultendpunct}{\mcitedefaultseppunct}\relax
\EndOfBibitem
\bibitem[Nair \latin{et~al.}(2008)Nair, Blake, Grigorenko, Novoselov, Booth,
  Stauber, Peres, and Geim]{Nair2008}
Nair,~R.~R.; Blake,~P.; Grigorenko,~A.~N.; Novoselov,~K.~S.; Booth,~T.~J.;
  Stauber,~T.; Peres,~N. M.~R.; Geim,~A.~K. \emph{Science} \textbf{2008},
  \emph{320}, 1308\relax
\mciteBstWouldAddEndPuncttrue
\mciteSetBstMidEndSepPunct{\mcitedefaultmidpunct}
{\mcitedefaultendpunct}{\mcitedefaultseppunct}\relax
\EndOfBibitem
\bibitem[Mak \latin{et~al.}(2008)Mak, Sfeir, Wu, Lui, Misewich, and
  Heinz]{Mak2008a}
Mak,~K.~F.; Sfeir,~M.~Y.; Wu,~Y.; Lui,~C.~H.; Misewich,~J.~A.; Heinz,~T.~F.
  \emph{Physical Review Letters} \textbf{2008}, \emph{101}, 196405\relax
\mciteBstWouldAddEndPuncttrue
\mciteSetBstMidEndSepPunct{\mcitedefaultmidpunct}
{\mcitedefaultendpunct}{\mcitedefaultseppunct}\relax
\EndOfBibitem
\bibitem[Li \latin{et~al.}(2008)Li, Henriksen, Jiang, Hao, Martin, Kim,
  Stormer, and Basov]{Li2008}
Li,~Z.~Q.; Henriksen,~E.~A.; Jiang,~Z.; Hao,~Z.; Martin,~M.~C.; Kim,~P.;
  Stormer,~H.~L.; Basov,~D.~N. \emph{Nature Physics} \textbf{2008}, \emph{4},
  532--535\relax
\mciteBstWouldAddEndPuncttrue
\mciteSetBstMidEndSepPunct{\mcitedefaultmidpunct}
{\mcitedefaultendpunct}{\mcitedefaultseppunct}\relax
\EndOfBibitem
\bibitem[Dawlaty \latin{et~al.}(2008)Dawlaty, Shivaraman, Strait, George,
  Chandrashekhar, Rana, Spencer, Veksler, and Chen]{Dawlaty2008}
Dawlaty,~J.~M.; Shivaraman,~S.; Strait,~J.; George,~P.; Chandrashekhar,~M.;
  Rana,~F.; Spencer,~M.~G.; Veksler,~D.; Chen,~Y. \emph{Applied Physics
  Letters} \textbf{2008}, \emph{93}, 131905\relax
\mciteBstWouldAddEndPuncttrue
\mciteSetBstMidEndSepPunct{\mcitedefaultmidpunct}
{\mcitedefaultendpunct}{\mcitedefaultseppunct}\relax
\EndOfBibitem
\bibitem[Yan \latin{et~al.}(2011)Yan, Xia, Zhu, Freitag, Dimitrakopoulos, Bol,
  Tulevski, and Avouris]{Yan2011}
Yan,~H.; Xia,~F.; Zhu,~W.; Freitag,~M.; Dimitrakopoulos,~C.; Bol,~A.~A.;
  Tulevski,~G.; Avouris,~P. \emph{ACS Nano} \textbf{2011}, \emph{5},
  9854--60\relax
\mciteBstWouldAddEndPuncttrue
\mciteSetBstMidEndSepPunct{\mcitedefaultmidpunct}
{\mcitedefaultendpunct}{\mcitedefaultseppunct}\relax
\EndOfBibitem
\bibitem[Scharf \latin{et~al.}(2013)Scharf, Perebeinos, Fabian, and
  Avouris]{Scharf2013}
Scharf,~B.; Perebeinos,~V.; Fabian,~J.; Avouris,~P. \emph{Physical Review B}
  \textbf{2013}, \emph{87}, 035414\relax
\mciteBstWouldAddEndPuncttrue
\mciteSetBstMidEndSepPunct{\mcitedefaultmidpunct}
{\mcitedefaultendpunct}{\mcitedefaultseppunct}\relax
\EndOfBibitem
\bibitem[Kliewer and Fuchs(1966)Kliewer, and Fuchs]{Kliewer1966}
Kliewer,~K.~L.; Fuchs,~R. \emph{Physical Review} \textbf{1966}, \emph{144},
  495--503\relax
\mciteBstWouldAddEndPuncttrue
\mciteSetBstMidEndSepPunct{\mcitedefaultmidpunct}
{\mcitedefaultendpunct}{\mcitedefaultseppunct}\relax
\EndOfBibitem
\bibitem[Hillenbrand \latin{et~al.}(2002)Hillenbrand, Taubner, and
  Keilmann]{Hillenbrand2002}
Hillenbrand,~R.; Taubner,~T.; Keilmann,~F. \emph{Nature} \textbf{2002},
  \emph{418}, 159--62\relax
\mciteBstWouldAddEndPuncttrue
\mciteSetBstMidEndSepPunct{\mcitedefaultmidpunct}
{\mcitedefaultendpunct}{\mcitedefaultseppunct}\relax
\EndOfBibitem
\bibitem[Zuev \latin{et~al.}(2009)Zuev, Chang, and Kim]{Zuev2009a}
Zuev,~Y.; Chang,~W.; Kim,~P. \emph{Physical Review Letters} \textbf{2009},
  \emph{102}, 096807\relax
\mciteBstWouldAddEndPuncttrue
\mciteSetBstMidEndSepPunct{\mcitedefaultmidpunct}
{\mcitedefaultendpunct}{\mcitedefaultseppunct}\relax
\EndOfBibitem
\bibitem[Wei \latin{et~al.}(2009)Wei, Bao, Pu, Lau, and Shi]{Wei2009}
Wei,~P.; Bao,~W.; Pu,~Y.; Lau,~C.; Shi,~J. \emph{Physical Review Letters}
  \textbf{2009}, \emph{102}, 166808\relax
\mciteBstWouldAddEndPuncttrue
\mciteSetBstMidEndSepPunct{\mcitedefaultmidpunct}
{\mcitedefaultendpunct}{\mcitedefaultseppunct}\relax
\EndOfBibitem
\bibitem[Falkovsky and Pershoguba(2007)Falkovsky, and
  Pershoguba]{Falkovsky2007}
Falkovsky,~L.; Pershoguba,~S. \emph{Physical Review B} \textbf{2007},
  \emph{76}, 153410\relax
\mciteBstWouldAddEndPuncttrue
\mciteSetBstMidEndSepPunct{\mcitedefaultmidpunct}
{\mcitedefaultendpunct}{\mcitedefaultseppunct}\relax
\EndOfBibitem
\bibitem[Falkovsky(2008)]{Falkovsky2008}
Falkovsky,~L.~A. \emph{Journal of Physics: Conference Series} \textbf{2008},
  \emph{129}, 012004\relax
\mciteBstWouldAddEndPuncttrue
\mciteSetBstMidEndSepPunct{\mcitedefaultmidpunct}
{\mcitedefaultendpunct}{\mcitedefaultseppunct}\relax
\EndOfBibitem
\bibitem[Wunsch \latin{et~al.}(2006)Wunsch, Stauber, Sols, and
  Guinea]{Wunsch2006}
Wunsch,~B.; Stauber,~T.; Sols,~F.; Guinea,~F. \emph{New Journal of Physics}
  \textbf{2006}, \emph{8}, 318--318\relax
\mciteBstWouldAddEndPuncttrue
\mciteSetBstMidEndSepPunct{\mcitedefaultmidpunct}
{\mcitedefaultendpunct}{\mcitedefaultseppunct}\relax
\EndOfBibitem
\bibitem[Hwang and {Das Sarma}(2007)Hwang, and {Das Sarma}]{Hwang2007a}
Hwang,~E.~H.; {Das Sarma},~S. \emph{Physical Review B} \textbf{2007},
  \emph{75}, 205418\relax
\mciteBstWouldAddEndPuncttrue
\mciteSetBstMidEndSepPunct{\mcitedefaultmidpunct}
{\mcitedefaultendpunct}{\mcitedefaultseppunct}\relax
\EndOfBibitem
\end{mcitethebibliography}
\end{document}